\documentclass[twocolumn,apjl]{openjournal}
\usepackage{graphicx}
\usepackage{amsmath}
\usepackage{amssymb}
\usepackage{xspace}
\usepackage[dvipsnames]{xcolor}
\usepackage[backref, pagebackref=false, hyperindex=true, breaklinks=true, colorlinks=true, urlcolor=magenta, linkcolor=NavyBlue, citecolor=cyan, citebordercolor={0 1 0}, pagecolor=red, bookmarks=true, filecolor=blue,bookmarksopen=true, plainpages=false, pdfpagemode=UseThumbs]{hyperref}
\hypersetup{
colorlinks=true,
linkcolor=blue,
filecolor=blue,
urlcolor=blue,
citecolor=blue,
}
\usepackage{orcidlink}
\usepackage{multirow}
\usepackage{soul}
\usepackage{multirow}

\newcommand{\sgn}{\mathop{\mathrm{sgn}}}
\begin{document}

\title{Detecting False Positives With Derived Planetary Parameters: Experimenting with the KEPLER Dataset\vspace{-16pt}}

\shorttitle{Kepler False Positives}

\shortauthors{Rafaih et al.}

\author{Ayan Bin Rafaih\orcidlink{0009-0005-8393-6756}}
\affiliation{Aitchison College, Lahore\vspace{-8pt}}

\author{Zachary Murray\orcidlink{0000-0002-8076-3854}}
\affiliation{Université Côte d'Azur, France\vspace{-8pt}}

\begin{abstract}
  Recent developments in computational power and machine learning techniques motivate their use in many different astrophysical research areas. Consequently, many machine learning models have been trained to classify exoplanet transit signals -  typically done by using time series light curve. In this work we attempt a different approach and try to improve the efficiency of these algorithms by fitting only derived planetary parameters, instead of full time-series light curves. We investigate and evaluate 4 models (Logistic Regression, Random Forest, Support Vector Machines and Convolutional Neural Networks) on the KEPLER dataset, using precision-recall trade-off and accuracy metrics. We show that this approach can identify up to $\approx 90\%$ of false positives, implying the planetary parameters encompass most of the relevant information contained in a light curve. Random Forest and Convolutional Neural Networks produce the highest accuracy and the best precision-recall trade-off. We also note that the accuracies as a function of the stellar eclipse flag (\texttt{SS}) have the best performance.
\end{abstract}

\keywords{Exoplanets --- Machine Learning --- Convolutional Neural Networks --- False Positives --- Logistic Regression --- Random Forest --- Support Vector Machines\vspace{12pt}}

\maketitle

\section{Introduction}
\label{sec:intro}

The first exoplanet was confirmed in 1992 by pulsar timing \citep{Wolszczan_1992}, this detection was followed shortly by  the first detection of an exoplanet around a main-sequence star in 1995 via radial velocities \citep{Mayor_1995}.  The first detected transit of an exoplanet was not until 2000, when HD 209458b was found to transit its star \citep{Charbonneau_2000}.

    Despite this rather late start, the transit method has since proved to be the most productive method of detecting new exoplanets, with more than five thousand confirmed planets found to date.  It is also the main method employed by large exoplanet searches, including the Kepler and TESS missions \citep{Koch_2010,Ricker_2015} and in addition to numerous ground based searches, including the WASP, TRAPPIST, KELT, NGST and many others \citep[e.g][]{Pepper_2007,Butters_2010,Jehin_2011,Wheatley_2018}.
    
    Large transit surveys typically search for transit-like features in their lightcurves using a box least squares algorithm \citep{Kovacs_2002}, or transit least squares algorithm \citep{Hippke_2019}.  However, not every candidate detected with these algorithms corresponds to an exoplanet. Sources of false positives include eclipsing binaries \citep{2003ApJ...593L.125B, 2006ApJ...651L..61O}, stellar contamination from blended sources, and instrumental artifacts \citep{Coughlin_2016}.

    False positives are often identified with additional spectroscopic or photometric observations. However, when such observations are not available, statistical methods are often used to estimate a false positive probability.  The early algorithms included the \texttt{VESPA} code \citep{Morton_2012,Morton_2015}, while more modern approaches include \texttt{TRICERATOPS} \citep{Giacalone_2021} or other machine learning algorithms \citep{Armstrong_2021,Malik_2022,TardugnoPoleo_2024}.
    
    In general, these false positive detection scripts operate on raw light curves to maximize the amount of available information for classification.  However, this approach has the downside of requiring rather sophisticated algorithms to work with the raw data.  In this work, we take a different approach.  We focus instead on the standard transit parameters (for example, transit depth and the impact parameter) computed for these false positives. Since the planetary parameters are essentially low-dimensional, yet physically motivated, summaries of the light curves, using them instead of the full light curves results in a significant simplification of the machine learning models.
    
    In this work, we investigate this trade off in accuracy.  We find algorithms that detect these false positives, as closely as possible, using only the Kepler derived planetary parameters.  We show that even simple regressions can separate the false positives from the true planets with accuracies exceeding $70 \%$, and more sophisticated models can do so with accuracies of $\approx 90\%$.  Finally we show that the models accuracies are maintained for certain types of contamination, especially those with features that are not-transit like or stellar eclipses, but do struggle for other types of systematics.

\section{Methods}
    For this study, we utilize data collected by the Kepler Space Telescope, launched by NASA in 2009 to survey a $115 \mathrm{deg^2}$ field of view of the northern sky for exoplanets \citep{Borucki_2010,Koch_2010}. Our training data is sourced from the Kepler Objects of Interest catalogue cumulative list, which is currently hosted at the NASA Exoplanet Archive.\footnote{\url{https://exoplanetarchive.ipac.caltech.edu/}} 
    On date of access, the data set contains a total of 9,564 entries and the individual characteristics of each body, including observed properties (e.g planet-star radius ratio) and derived ones (e.g. the effective temperature).
    The Kepler catalogue also provides a ternary classification feature, derived form a numerical disposition score.  This score is a numerical value that expresses the confidence in the category classification and  is calculated using a Monte Carlo algorithm that equates the score's value to the proportion of iterations where the Kepler Robovetter algorithm \citep{2017ksci.rept...22C, 2018ApJS..235...38T,2020ascl.soft12006C}  classifies each threshold-crossing event as a potential candidate. 
    
    The catalogue consists of three types of entries: (i) confirmed exoplanets using the NASA exoplanet archive disposition, (ii) candidate exoplanets using the Kepler Catalogue Disposition, and (iii) False Positive exoplanets. For candidates, a higher score represents a greater amount of confidence in its disposition, while for false positives, a higher value indicates less confidence in its classification. For entries whose official disposition is `candidate', a higher score represents a greater amount of confidence in its disposition while for entries whose disposition is `false positive', a higher score represents less confidence in its disposition (and thus there is a greater chance of it being a candidate). For our study, we choose a binary Kepler disposition feature list of exoplanets as either `Candidates' or `False Positives', from the ternary set in the archive. There are in total 9,564 entries in our extracted binary catalogue of which 4,717 are candidates and 4,847 are false positives. 

\begin{figure}
        \centering
        \includegraphics[width=0.48\textwidth]{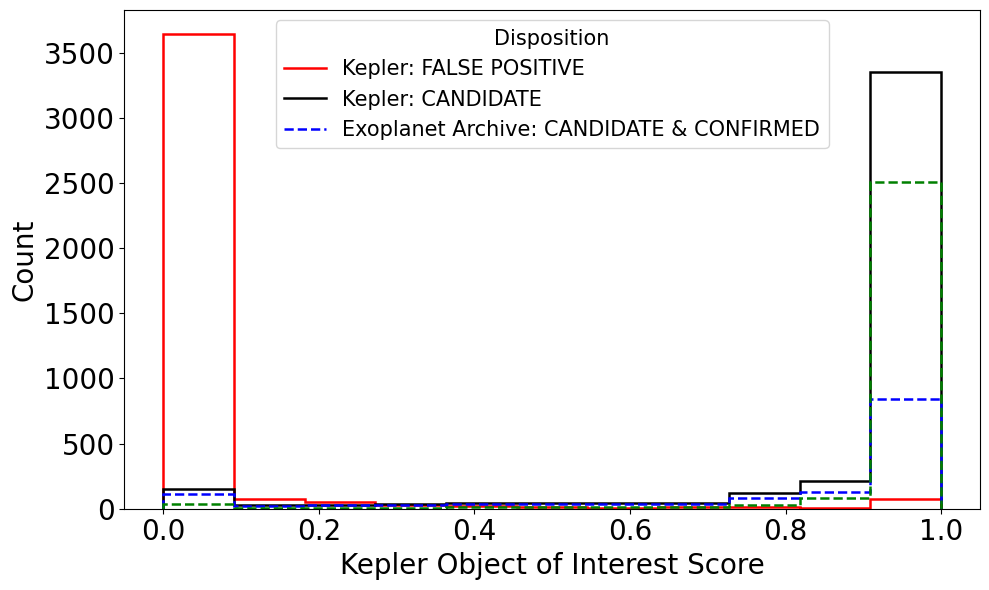}
        \caption{We show the distribution of $\sigma$ (koi\_score) across the categories provided in the disposition. We note that the false positives remain the same in number, with high confidence in their dispositions. On the other hand, most of the candidates either preserve their vetting status, or become confirmed exoplanets through dispositional vetting methods employed outside of the Kepler data. There's a very small, thin spread of candidates across the entire range of koi\_score, which is however insignificant. For example, we note that there are 2,736 candidates that changed their vetting status to "Confirmed", while the remaining 1,981 candidates most have a koi\_score $\sigma > 0.80$.}
        \label{fig:histogram}
    \end{figure}

    As shown in Figure \ref{fig:histogram}, the distribution of scores for both the categories follows a sharply peaked distribution: the false positives are concentrated towards score of $\sigma = 0$, while the candidates are concentrated towards $\sigma = 1$. A large proportion of the candidates planets are converted to confirmed exoplanets, while some retain their Kepler vetting status. Since the number of candidates with $\sigma < 0.2$ is very low, we adopt the binary categories provided by Kepler disposition. Additionally, we can also see the relatively concentrated spread of the candidates that don't become confirmed exoplanets, near the $\sigma > 0.80$ threshold, showing the high confidence in the remaining planets.

\begin{table*}[htbp]
\centering
\caption{Feature list used for training,testing and evaluating the models. Features marked with * are excluded from training/testing but retained for comparison.}
\begin{tabular}{|l|l|}
\hline
\textbf{Feature Name} & \textbf{Description} \\
\hline
koi\_pdisposition & Classification label for the exoplanet entry; either \texttt{CANDIDATE} or \texttt{FALSE POSITIVE} based \\
                  & on Kepler's disposition.\\
koi\_pflag\_nt* & Not-transit-like flag; for entries whose light curves don't resemble typical planetary\\
               & transits (e.g., variable stars).\\
koi\_pflag\_ss* & Stellar eclipse flag; indicates eclipsing binary systems with significant  \\
               & secondary eclipses or variability. \\ 
koi pflag\_co* & Centroid offset flag; marks cases where the transit signal is offset from the target star. \\
koi pflag\_ec* & Ephemeris match contamination flag; identifies signals that match known periods/epochs of\\
              & other objects, suggesting contamination. \\
koi\_period (days) & Orbital period; time between successive transits. \\
koi\_impact & Impact parameter; normalized distance between planet and stellar center during mid-transit. \\
koi\_duration (hours) & Transit duration \citep{2002ApJ...580L.171M}. \\
koi\_depth (ppm) & Transit depth; fractional drop in stellar brightness during transit. \\
koi\_sma* (AU) & Semi-major axis of the orbit. \\
koi\_model\_snr & Transit signal-to-noise ratio. \\
koi\_bin\_oedp\_sig & Odd-even depth comparison statistic. \\
koi\_steff (K) & Stellar effective temperature. \\
koi\_srad (Solar radii) & Stellar radius. \\
\hline
\end{tabular}
\label{tab:featurelist}
\end{table*}

\begin{figure}[h!]
    \centering
    \includegraphics[width=0.38\textwidth]{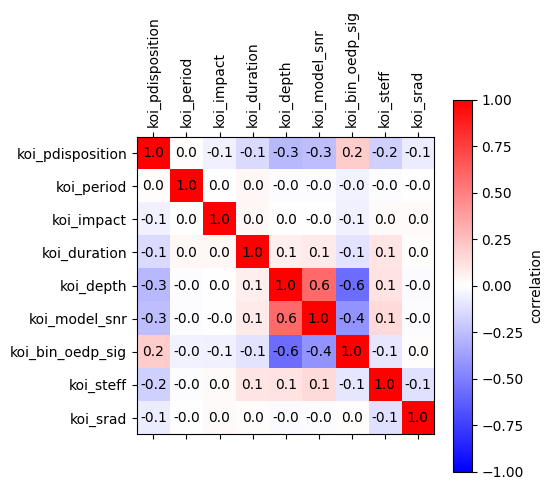}
    \vspace{1em}
    \includegraphics[width=0.48\textwidth]{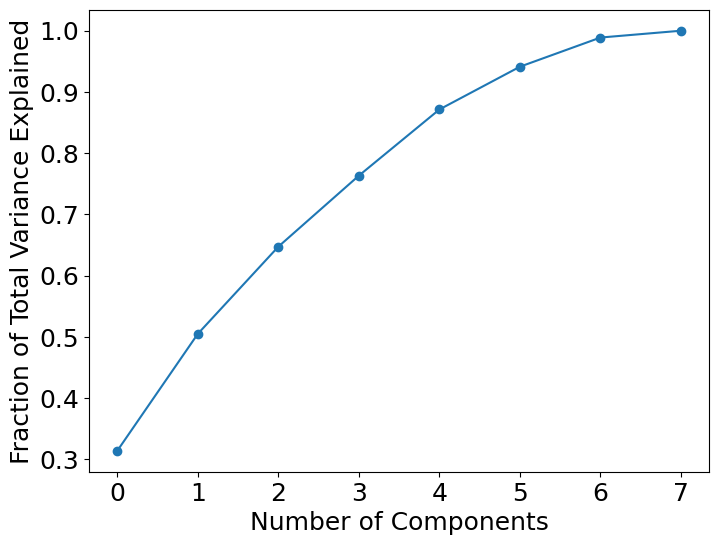}
    \caption{Top: Correlation matrix of the features. Bottom: Cumulative variance explained by principal components. Adjacent to the diagonal of perfect correlations, we can see a concentrated region that exhibits relatively significant correlations between three of the features: transit depth, transit signal-to-noise ratio and the odd-even depth comparison statistic. Across the upper and left edge, there's a relatively weak region of correlation with the target disposition feature. (bottom) The cumulative variance plot shows the total variance provided by the first $N$ principal components. Since the curve doesn't fully plateau to a constant variance till the 8th component, we can conclude that each of the components provides some level of variance to the dataset and can't be fully discarded.}
    \label{fig:correlation_pca}
\end{figure}

\subsection{EDA and Data Prep} 
\label{EDA}
    In machine learning, it is typical to choose features that maximize the performance of the algorithm. When choosing these features we prioritize those extracted directly from the light curves to minimize the effects of additional assumptions on our algorithmic performance, as shown in table \ref{tab:featurelist}. For example, we use the parameters \texttt{koi\_depth} and \texttt{koi\_model\_snr} instead of \texttt{koi\_stellar\_magnitude}, since the \texttt{koi\_model\_snr} can be determined independently from an uncalibrated light curve, without a reference magnitude. This results in a set of initial features that can all be derived from a photometric measurements using an uncalibrated light curve and stellar spectrum (e.g. \texttt{koi\_steff} and \texttt{koi\_srad}). A complete summary of the features identified is given in Table \ref{tab:featurelist}.    

    Having selected a suite of potentially important features, we then aim to eliminate any redundant or degenerate parameters. To verify if all our parameters are necessary, we examine our dataset using Principal Component Analysis (PCA). PCA reduces the dimensions of a $N$-dimensional data into a $M$-dimensional data where $N>M$ by constructing orthogonal eigenvectors (principal components) preserving the maximum variance $\gamma$ in each one of them. This is done by computing a covariance matrix for the dataset, whose eigenvalues and eigenvectors are fit to produce a feature vector with ranked variance for each of the ordered principal components. If all the principal components contribute significantly to the variance, then it implies they are all useful for predicting a planet's disposition status. However, if one or more components were found to not contribute significantly to the variance of the features, we might drop those components to simplify the problem. In Figure \ref{fig:correlation_pca}, we show the cumulative explained variance by the PCA. Since it flattens only slightly as we move towards higher components, we conclude that all of our 9 chosen features from the principal components are necessary for the problem.
    
    Therefore, we selected a total of 13 columns, dropping \texttt{koi\_pflag\_nt}, \texttt{koi\_pflag\_ss}, \texttt{koi\_pflag\_co} and \texttt{koi\_pflag\_ec} from the catalogue, out of which 9 were used for training and testing our models and 4 were used for prediction comparisons. We produce a correlation matrix of the features to determine the relationship between the individual parameters, as shown in Figure \ref{fig:correlation_pca}. This was done to ensure the importance of each feature for the model's performance and to effectively remove any derived parameters. We notice small values for our correlation, indicating that each of the feature selected is largely independent of the others. Furthermore, we also make use of the 4 flag parameters in the table. These are features that categorise a given Kepler Object of Interest (KOI) into one of four possible classifications. We use these parameters to validate the model's capabilities in detecting different types of KOIs.

    We binarize the exoplanet disposition and remove all rows with missing or null values, giving us 7,995 entries in total. Specifically, the new dataset includes 3,901 false positives and 4,094 candidates. The datapoints are particularly concentrated in certain regions of the feature space when plotted out in their raw form. To help separate out densely packed points, the feature matrix $\mathbf{X}$ was scaled according to the function $\mathbf{X}_{new} = \log_{10}({\mathbf{X}_{old}}+2)$. The logarithmic function improves the distribution of the data for all the features , allowing for better visualisation and quicker algorithmic convergence. Additionally, the function also includes an offset of 2 with $\mathbf{X_{old}}$ since the minimum of all the essential parameters chosen above is -1, for example \texttt{koi\_bin\_oedp\_sig}. This offset ensures that the logarithm is always a real number.

    The dataset is split into two parts: (i) 70\% of it is reserved for training the models and (ii) 30\% of it is used for the testing set. For each of the divisions and models, the splits are randomized without replacement and the individual datasets are standardised to between 0 and 1.  The data is resampled using an oversampler that shuffles through the dataset randomly. This is only done for the training section of the dataset, and not done for the testing fragments. For each feature in the dataset, the standard deviation $\sigma$ and mean $\mu$ are calculated. Each column is then sigma clipped, removing any row with an entry more than 5 standard deviations from the mean. This is done to ensure all the data lies close to each other, and any outliers, which might distort robust classification of the transit signal, are removed.
    
    \begin{figure*}[!ht]
    \centering
    \includegraphics[width=1.0\textwidth]{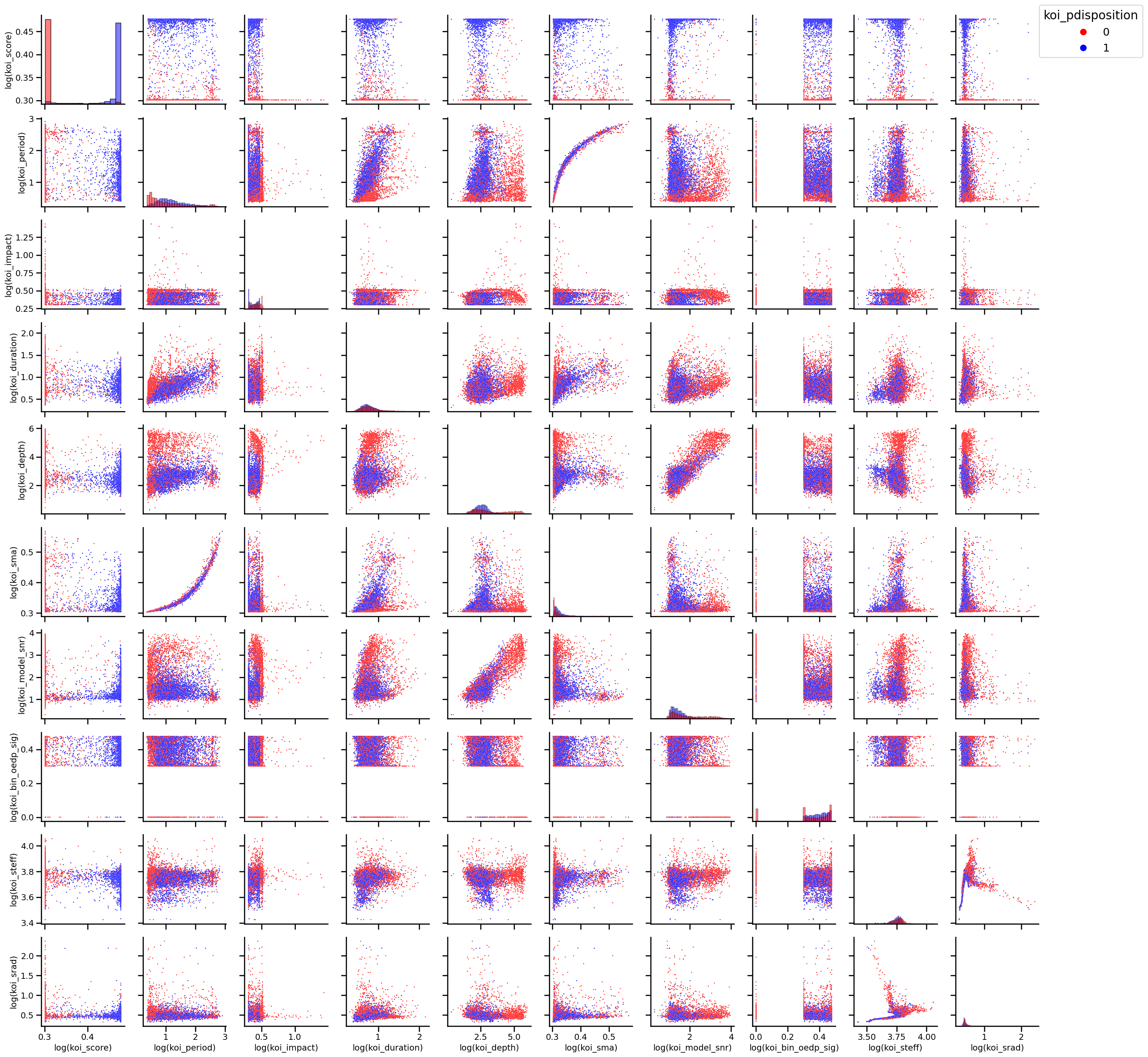}
    \caption{Pairplots for the training set with logistic regression ground-truth data points. The plots are color-coded as follows: red plots represent false positives and blue plots represent candidates. Some of the subplots (\texttt{koi\_model\_snr} vs \texttt{koi\_depth}, \texttt{koi\_duration} vs \texttt{koi\_model\_snr}) show a clear, discrete difference in distribution between each of the plots since the logistic regression model is able to separate the features effectively and cluster the training predictions in separate regions. For example for \texttt{koi\_model\_snr} vs \texttt{koi\_depth}, the model classifies signals with a higher transit depth and a higher odd-even comparison statistic as false positives, as the two parameters are intrinsically linked to the same transit properties since the transit-signal-to noise ratio is represented as a standard value, calculated by taking the average of the mean flux measurements. Note that we only show \texttt{koi\_sma} for its direct relationship to \texttt{koi\_period}, due to Kepler's law, hence it's not included for the actual training set since it's just a degenerate parameter. We also see a very small, concentrated distribution of the stellar radii in the \texttt{koi\_srad} graph.}
    \label{LogisticRegression}
    \end{figure*}

\subsection{A simple model: Logistic Regression}
\label{logisticsectionref}
    Once preliminary data preparation is done, we begin by fitting a simple model to the data to serve as a baseline to which we compare our more sophisticated machine learning techniques. Since we predict over a categorical variable, a logistic regression is ideal for this task.
    Logistic regression models the probability of a given classification by using a linear predictor function and regression coefficients.  The model is defined as:

    \begin{eqnarray}
        \hat{y} = \frac{1}{1 + e^{-(\mathbf{X} \cdot \vec{w} + b)}}
    \end{eqnarray}
    where $\vec{w}$ is the vector of weights, and $b$ is the offset. 
    We solved for the weights using \texttt{SKLEARN's} logistic regressor, where we specified 10,000 iterations to be the limit for consideration of an optimal converging solution \citep{scikit-learn}. 
    
    There are some observations worth note. For example, the plots for \texttt{koi\_srad} with the transit depth, duration and the transit signal-to-noise ratio are more concentrated near to the y-axis, with less scattered points around the median. Another notable observation is the graph for \texttt{koi\_steff} vs \texttt{koi\_srad}, which resembles a standard Hertzsprung-Russell diagram, in which most of the false positives are distributed in the giants branch. 
    Additionally, for \texttt{koi\_bin\_oedp\_sig}, we can observe a relatively higher number of candidate exoplanets, dominating the vector space, with the false positives at the edges. A non-significant cluster division is only observed with \texttt{koi\_model\_snr}, since the two relate to each other: \texttt{koi\_model\_snr} represents the transit signal to noise ratio which has been from normalised transit depth, and then \texttt{koi\_bin\_oedp\_sig} is a comparison statistic for the type of the transit depth. 
    In total, we show 5596 training probabilities and discrete predictions, and similarly 2399 testing probabilities and discrete predictions.

    \begin{figure}[h!]
    \centering
    \includegraphics[width=0.47\textwidth]{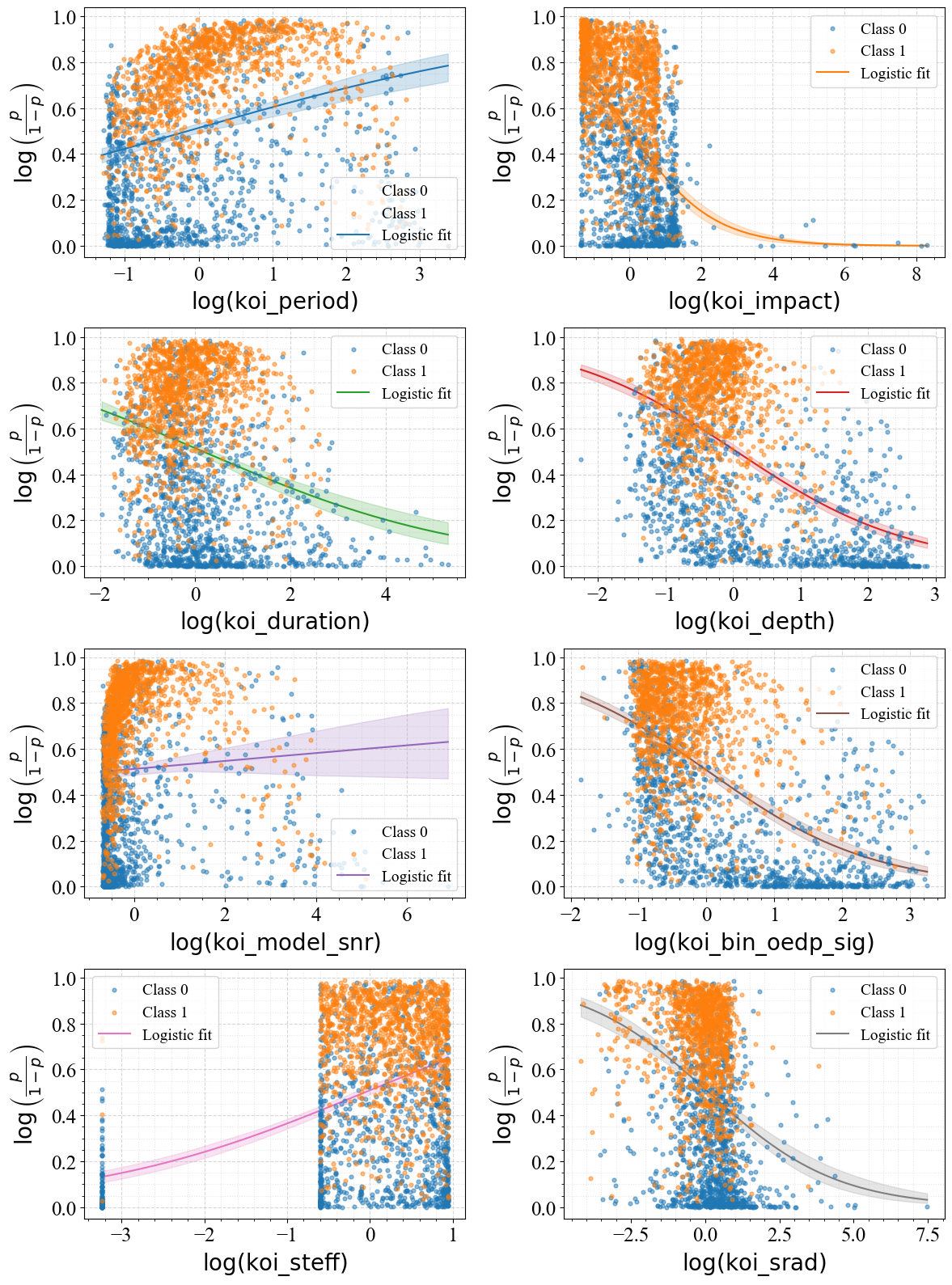}
    \caption{We use a standard logit transformation, defined by $\log{(\frac{p}{1-p})}$, on the y-axis for each of them to help linearise the model's decision function. The spread and distribution of the continuous probability predictions from the Logistic Regression model are shown here in the parameter space, which are made through 100 bootstrap iteration, which are all the same size as the training set, each one involving an individual logistic fit. The logistic plots show a varied, non-linear distribution of the predictions across the parameter space, which is more consistent with the nature of the features and their relative correlations.}
    \label{logitoverallplot}
    \end{figure}
    
    While the default approach for binary classification problems is to assume a threshold of 0.5 as a cut-off between classes, it may not be optimal for every use-case.  Therefore, we evaluate the metrics accuracy, precision, recall and $F_1$ score on various thresholds between 0 and 1 at increments of 0.01, to provide a better idea of metric trade-offs at various thresholds. We can define our metrics for precision, (P), recall (R) and $F_1$ score ($F_1$) as follows:
    \begin{equation}
    \label{eq(2)}
        P = \frac{TP}{TP+FP}
    \end{equation}
    \begin{equation}
    \label{eq(3)}
        R = \frac{TP}{TP+FN}
    \end{equation}
    \begin{equation}
    \label{eq(4)}
        F_1 = 2 \cdot \frac{P R}{P+R}
    \end{equation}

    where $TP$ is the number of true positives, $TN$ the number of true negatives, $FP$ the number of false positives and $FN$ the number of false negatives.     
    To quantify the accuracy threshold trade-off, we use the logarithmic offset dataframe which gives us a smooth curve that rises to a local maxima at a threshold less than 0.50, after which it decreases to a constant level at thresholds very close to 1.0.
    The precision recall (PR-AUC) curves have an area under the curve $P_{AUC} = 0.75$, signifying that the model is unable to differentiate between the type of transit signals at certain threshold values, struggling to single out the important feature vectors. The varying decision thresholds affecting the recall do make a good case for not choosing the conventional 0.5 as the threshold for the dataset to maintain consistency and good performance on a wide variety of datasets, thereby reducing bias towards the specific set of parameters in play. 
    Other decision thresholds favour particular metrics for the model. For example, choosing a threshold in the subsets $\epsilon \approx 0.25$ or $\epsilon \approx 0.80$ maximises $F_{1}$-score and precision respectively.
    
\begin{figure}[!ht]
    \centering
    \includegraphics[width=0.48\textwidth]{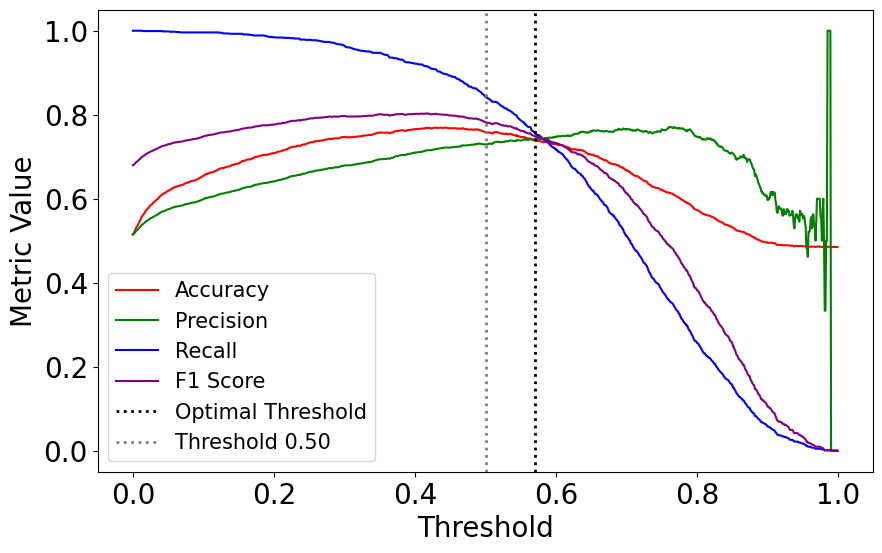}
    \caption{Plot for the Metrics against the threshold that differentiates between the false positive transit signals and the candidates transit signals. There are two dotted lines for thresholds of note: one on the conventional 0.50 where the model seems to sacrifice precision for a greater accuracy score. The 0.50 line intersects them at the point, where the $F_{1}$-score is above the accuracy and precision and the recall being a local maxima at that threshold. The second dotted line presents a first case for an optimal threshold where we consider the algebraic case defined in section \ref{logisticsectionref}, when precisions equals recall. Therefore, it represents a possible intersection points for the precision, recall and $F_{1}$ score metrics. The accuracy metrics, due to its nature, doesn't intersect at a common point, which could be denoted as the balance between all 4, which is simply not possible, both algebraically and practically.}
    \label{thresholds}
    \end{figure}

    An optimal threshold could be identified if all of the three metrics are proportionately important in a binary classification problem: This may represent an intersection point for all the curves where the performance is balanced across all of the defined metrics. However, as per the need of the problem, we could favour one metric over another, such as precision over the recall. For example, if $P=R$, then we can confidently say that there exists a point such that $F_{1}=P=R$. We iterate through each one of the metrics as a index-based loop and calculate the range difference as the maximum of the metrics - minimum of the metrics. Then, we find the minimum possible argument value for our range different, which helps us to approximate the a trade-off threshold as shown in Figure \ref{thresholds}, where the curves for precision, recall and $F_{1}$ score intersect each other. However, the final decision threshold to act as the contrast line between false positives and candidates depends on the requirements of the system. 
    As our base model, Logistic Regression is able to separate some of the primary features on the parameter space, but does not help to differentiate between the two types of exoplanet transit signals any better above a set boundary, which means more sophisticated models would be needed to help separate out the parameters better, yielding higher accuracies for both types of classes.

\subsection{Random Forest}
    
    Unlike logistic regression, random forest classification works by combining trees which iteratively split the predictors. Since these splits are binary, random forest can rapidly represent more complex behaviour than can't be captured with a logistic model. To investigate whether there are additional details not captured by a logistic regression we fit a random forest model using the Random Forest Classifier (RFC) in \texttt{SKLEARN} over our same logarithmic dataset. We use a train-test split of $70:30$ for all our random forest models.  
    Hyper-parameter tuning is carried out to optimise the model of the performance, using a mixture of exhaustive and case-specific tuning. 
    
    A parameter grid is defined to help tune all the important hyper-parameters for the models with wide range of possible cases, as defined in table \ref{tab: hyperparameterlist}. If we were to explore all these possibilities with \texttt{GridSearchCV} - with a 5 fold cross-validation for each of the all possible candidates - it would result in up to 20 million fits, which was computationally prohibitive. Instead, we use \texttt{RandomSearchCV} to randomly sample a subset of the combinatorial hyper-parameter grid.  We sample 2000 random samples in the hyper-parameter grid, and fit using 5 fold cross-validation, resulting in 10,000 fits. Considering $N$ folds at the start for our algorithm, these are inputted as the number of splits needed for the K-Fold Validation that we implement before running the classifier on the parameters. We randomise the dataset and shuffle the rows for each of the folds, yielding $\frac{T_{y}}{N}$ splices, where $T_{y}$ represents the total number of rows in the training and testing dataset. Each splice contains correct and incorrect predictions for the binary class problem in this paper, from which an accuracy score is derived for each of the folds using cross-validation score. Each score is treated as a separate score and a mean average final accuracy is used for analysis. We select the ten models with the highest mean accuracies. From here the hyper-parameters were manually refined, by prioritizing those that appeared frequently in this grid search. For example, for our \texttt{num\_estimators} hyper-parameter, we initially had 5 possible values, as shown in table \ref{tab: hyperparameterlist}. Of those, only 3 appear among the highest mean accuracies and only those were preserved for the next iteration.
    
    After identifying the most promising hyper-parameters, we then repeat our random search, with cross validation, over the combinatoric grid of the remaining hyper-parameters. 
    Since many of the combinations have been discarded due to their adverse affect on model performance, the number of remaining combinations feasibly allow for a \texttt{GridSearchCV}. Therefore, an exhaustive search is done on the remaining parameters where 5 folds are fitted for each of the 1,536 candidates, totalling 7,680 fits. We report the model with the highest accuracy of this search. This model has the following tuned hyper-parameters: 300 estimator trees, entropy criterion, tree depth of 20, $\sqrt{N}$ features looked at (where $N$ were the total number of features), minimum sample split of 10, at least 1 sample for a leaf, with no restrictions on \texttt{class\_weight}, \texttt{max\_leaf\_nodes} and \texttt{max\_samples}.

    To judge the number of correct and wrong predictions for both the classes we make a receiver operating characteristic (ROC) curve to compare the true positive rate with the false positive rate. The ROC curve yields an area of $1.00$ for the training predictions and an area of $0.94$ for the testing predictions.
   
    \begin{table*}[ht]
    \centering
    \footnotesize
    \begin{tabular}{|c|c|c|}
        \hline
        \textbf{Hyper-parameter} & \textbf{Parameter Description} & \textbf{Parameter Grid} \\ 
        \hline
        n\_estimators & Total Number of trees in the Forest & \{100, 200, 300, 400, 500\} \\ 
        max\_depth & Maximum tree depth & \{None, 10, 20, 30, 40\} \\
        min\_samples\_split & Minimum samples required to split an internal node & \{2, 5, 10\} \\
        min\_samples\_leaf & Minimum samples at a leaf node for both branches & \{1, 2, 4\} \\
        max\_features & Number of features considered for splitting a node & \{sqrt, log2, None\} \\
        bootstrap & Whether bootstrap samples are used & \{True, False\} \\
        criterion & Function to measure split quality & \{gini, entropy, log\_loss\} \\
        max\_leaf\_nodes & Maximum leaf nodes in the tree & \{None, 50, 100\} \\
        min\_impurity\_decrease & Impurity threshold for node splitting & \{0.0, 0.01, 0.1\} \\
        min\_weight\_fraction\_leaf & Minimum weighted proportion of weights at a leaf node & \{0.0, 0.1, 0.2\} \\
        ccp\_alpha & Minimal cost-complexity pruning parameter & \{0.0, 0.01, 0.1\} \\
        max\_samples & Samples drawn for training each estimator & \{None, 0.5, 0.7, 0.9\} \\
        class\_weight & Weights assigned to each class & \{None, balanced, balanced\_subsample\} \\
        \hline
    \end{tabular}
    \caption{Hyper-parameter grid used to optimise model performance. Each of the hyper-parameters was used in the first iteration of the random search sampling. After the first iteration, \texttt{min\_weight\_fraction\_leaf} and \texttt{ccp\_alpha} are removed from the grid, while for others the number of combinations are reduced to the most important hyper-parameters only.} 
    \label{tab: hyperparameterlist}
\end{table*}

\subsection{SVM}
    Support Vector Machines (SVM) find the optimal hyperplane defined by $\Pi_0 := \vec{w}^T\textbf{X} + b = 0$ where $b$ is the intercept and $\vec{w}$ are the weights. The weights are chosen to best separate the binary labels for a dataset by maximizing the margin between them. The hyperplane $\Pi_{0}$ is separated using the boundary conditions for being greater or less than 0, if $y_{j}$ is 1 or -1, respectively. 
    The decision rule classifies a new observation by computing 
    \begin{equation}
        F(x_j)= \vec{w}^T x_j + b 
    \end{equation}
    with the observation assigned to $y_j = \sgn(f(x_j))$ where the defined marginal classifier is $y_{j}(b+ \vec{w}^T x_j) \geq (1-\epsilon_{j{}})$, with $\epsilon_{j}$ representing slack variables to help determine the side on the hyperplane on which the $j^{th}$ observation would be found. 

    We use the \texttt{SKLEARN}'s implementation of Support Vector Classifier (SVC) for our preliminary analysis of the classification problem. The metrics obtained from this classification form our basis value against which all the other variations for the SVM are compared against, which are discussed in the results section. We tune the hyper-parameters for the SVM, including the \texttt{kernel} and number of cross-validation folds, which are done through a manual selection, and an iterative approach, respectively. For kernels, we use the hyper-parameter grid \texttt{\{'rbf', 'linear', 'poly'\}}, where the degree of the polynomial varies from 0 to 9.  In addition, we examine cross validation folds from 2 to 10. 
    Therefore, in total we do a total of $12*9 = 108$ fits. With each iteration for the folds and the polynomial kernels in the SVC, we also make ROC-AUC curves to help us evaluate the model's performance with different parameters. 
    
\subsection{CNNs}
\label{cnn_explanation}
    Convolutional Neural Networks (CNN) \citep{abadi2016tensorflowlargescalemachinelearning} operate by approximating a function by sequentially convolving a set of chosen basis functions. They differ from RFC in that there are no discrete decision boundaries and they differ from SVM in that they are more flexible in terms of the output predictions. Let $\vec{\varphi} \in \mathbb{R}^{N}$ represent a vector $\vec{\varphi}$ in the set of all real numbers with $N$ elements. Extending the same convention, we can thus define a large matrix $\boldsymbol{\varphi}$ with $\phi_{1}$ rows and $\phi_{2}$ columns. The CNN takes the matrix as an input, which is sequentially fed through the convolutional layer in a forward push. This is done continuously till the last processing layer, yielding the desired classification for the 2 categories.
    
    We evaluate the dataset on three different architectures of neural networks using a sequential model, which arranges all of our layers in a linear stack. Model $M_{1}$ is simpler in architecture compared to model $M_{2}$ and $M_{3}$ in terms of layer structure and number of convolutions. 
    
    For each of the models, we choose a batch size of 50 and train the model for 500 epochs.  We reshape our feature matrix into a one dimensional vector before passing it into our models. Additionally, we also use binary cross-entropy as the loss function, Adam optimiser and the accuracy metric \citep{kingma2017adammethodstochasticoptimization}.
    
    For $M_{1}$, we used 3 convolutional blocks followed by a global pooling layer and a classification head. The global pooling layer computes the mean of each map across the input and regularises the network to identify globally important features. The classification head consists of a single neuron that aggregates the pooled features for the binary predictions. Finally, the sigmoid activation layer at the end converts the final output in the range $[0,1]$, representing the probability of the correct predictions for each class. Each of the convolutional blocks uses the same padding to preserve the dimensions of the input and also includes a \texttt{Conv1D} layer with the \texttt{ReLU} activation function. 
    
    At the end of the block, there is a batch normalisation layer which normalises the activations of the previous layer. In total, $M_{1}$ utilises $200,321$ total parameters, out of which $199,297$ parameters are trainable. 

    Models $M_{2}$ and $M_{3}$ have a more complex architecture than $M_1$, and consist of more layers.  We use a complicated architecture to create a wider layer for feature combination. 
    These two models are similar to each other except for two hyper-parameters: (i) $M_{2}$ uses hyperbolic \texttt{tanh} and \texttt{swish} activation functions instead of a uniform \texttt{ReLU} approach as in $M_{3}$, and (ii) $M_{2}$ makes use of the \texttt{Conv1D} layers instead of a 2D approach followed in $M_{3}$. 
    
    The new architecture consists of six convolutional blocks with batch normalisation and dropout regularisation to prevent over-fitting to the dataset. Of these blocks, the \texttt{tanh} activation function is used in the first, second and fifth convolutional blocks while the others use the \texttt{swish} function. We also include a dropout regularisation parameter of $30\%$ in the second, fourth and sixth convolutional blocks.  The number of neurons decrease by a factor of two, with each new dense layer, starting from 512 and going till 128, finally reaching the single neuron with sigmoid.
    
    Otherwise, both models follow a similar structure to $M_1$ in that they use a global average pooling layer and four dense layers. Both the models use $1,689,153$ total parameters out of which $1,686,465$ are trainable. The additional model details can be seen in table \ref{tab:M3details}.
    
    \begin{table*}[htbp]
    \centering
    \footnotesize
    \begin{tabular}{lccccc}
    \hline
    Layer & Output Size (LxS) & Activation & Normalization & Pooling & Dropout \\
    \hline
    Conv2D (8,1) & Lx1x256 & ReLU & Batch Norm & - & - \\
    Conv2D (5,1) & Lx1x512 & ReLU & Batch Norm & - & - \\
    Dropout & Lx1x512 & - & - & - & 30\% \\
    Conv2D (5,1) & Lx1x256 & ReLU & Batch Norm & - & - \\
    Conv2D (3,1) & Lx1x128 & ReLU & Batch Norm & - & - \\
    Dropout & Lx1x128 & - & - & - & 30\% \\
    Conv2D (3,1) & Lx1x128 & ReLU & Batch Norm & - & - \\
    Conv2D (3,1) & Lx1x64 & ReLU & Batch Norm & - & - \\
    Dropout & Lx1x64 & - & - & - & 30\% \\
    GlobalAvgPool & 64 & - & - & Global Avg & - \\
    Dense & 512 & ReLU & - & - & - \\
    Dropout & 512 & - & - & - & 50\% \\
    Dense & 256 & ReLU & - & - & - \\
    Dropout & 256 & - & - & - & 50\% \\
    Dense & 128 & ReLU & - & - & - \\
    Dropout & 128 & - & - & - & 50\% \\
    Dense & 1 & Sigmoid & - & - & - \\
    \hline
    \end{tabular}
    \caption{This table shows the neural network architecture for Model $M_{3}$, which is also quite similar to Model $M_{2}$.}
    \label{tab:M3details}
\end{table*}

    \begin{figure}[!ht]
    \centering
    \includegraphics[width=0.48\textwidth]{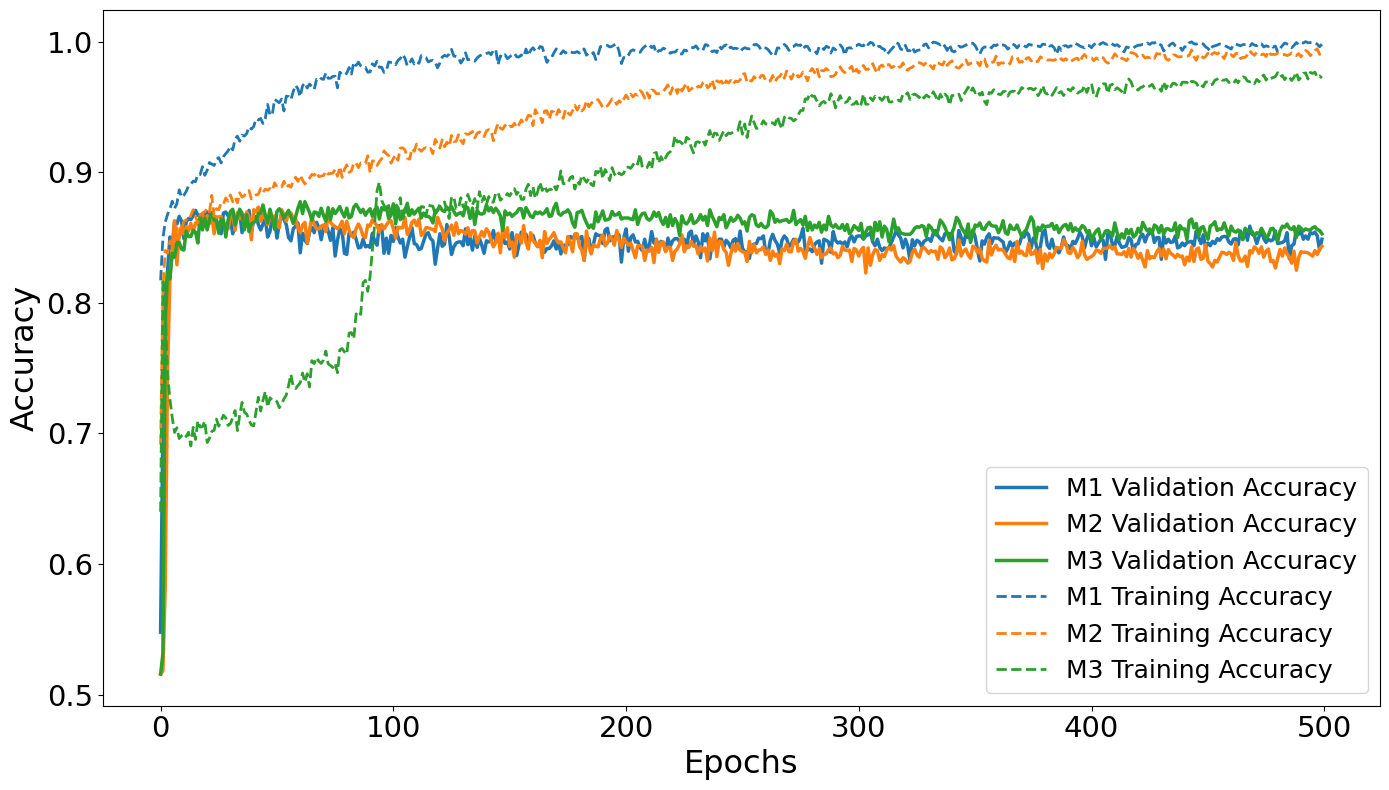}
    \caption{This plot shows the respective analysis of the performances of all the CNN architectures discussed above in section \ref{cnn_explanation} for $M_{1}$, $M_{2}$ and $M_{3}$. We can observe the validation accuracy for all the model architectures to plateau and reach a relatively constant accuracy of $\approx 87\%$. The consistency observed in all of the model architectures seen helps us conclude that our architectures are robust in the classification of exoplanets. 
    However, the best accuracies - and those models which do not overfit - occur at smaller epochs. Consequently we use a early stopping criterion for each model: terminating $M_{1}$ at epoch 32, $M_{2}$ at epoch 37 and $M_{3}$ at epoch 209.}
    \label{cnn_loss_and_accuracy_plot}
\end{figure}

    For all our models, we plot Figure \ref{cnn_loss_and_accuracy_plot}, which shows the accuracy and validation accuracy.    

\section{Results}
\label{sec:final_results}
    To help evaluate the performance of the models, we present two metrics, the train and validation accuracies, as they can be applied to each model regardless of the architecture. Each of the metrics are written with their standard deviations to represent the uncertainty in the data value. All of these are shown in Table \ref{tab:model_comparison3}.  For all models except CNNs, the standard deviations for the final value is calculated using $N_{0}$ bootstrap iterations, where $N_{0}$ is the length of the split containing the true classifications, individually fitting the model and taking the standard deviation over all trials. For CNNs, we notice the size of the fluctuations in accuracy are relatively constant as a function of the epoch.  Hence, we estimate  uncertainty by computing the standard deviation over the epochs 400-500. Choosing the last epochs helps to eliminate any fluctuations at the beginning of training, for example in model $M_{3}$. For SVMs, we utilise the cross-validation score over our defined fits to product a mean accuracy for each model variation. 
    
    Our results suggests that there is an upper bound to the best model performance that can be obtained by using only the derived planetary parameters instead of extracting data points from light curves, which we show in table \ref{tab:model_comparison3}, as the validation accuracy for our more sophisticated models approaches this upper limit of $\approx 92.2\%$. Additionally, we can clearly see that our simpler models such as logistic regression and SVMs, underperform compared to random forest and CNNs despite hyper-parameter tuning. This is partly because of the simplicity of the model which leads to it being unable to capture the full information in the parameters fed into it. Other notable observations are the significantly higher standard deviations observed for the accuracies of the CNNs compared to the other models.

    \begin{table}[!htbp]
    \centering
    \begin{tabular}{lcc}
    \hline
    \hline
    Model & Train Accuracy & Validation Accuracy \\
    & (\%) & (\%) \\
    \hline
    Logistic Regression & $76.516 \pm 0.602$ & $77.544 \pm 0.735$ \\
    Random Forest & $96.590 \pm 0.236$ & $87.803 \pm 0.574$ \\
    SVM (RBF) & $83.485 \pm 0.002$ & $83.403 \pm 0.010$ \\
    SVM (Linear) & $77.434 \pm 0.003$ & $77.241 \pm 0.015$ \\
    SVM ($9$ Folds) & $83.554 \pm 0.002$ & $83.441 \pm 0.015$ \\
    SVM (Poly) & $86.589 \pm 0.003$ & $86.205 \pm 0.009$ \\
    CNN $M_{1}$ & $91.99 \pm 0.284$ & $88.591 \pm 0.687$ \\
    CNN $M_{2}$ & $87.87 \pm 0.181$ & $89.023 \pm 0.466$ \\
    CNN $M_{3}$ & $92.14 \pm 0.343$ & $\boldsymbol{92.233 \pm 0.335}$ \\
    \hline
    \end{tabular}
    \caption{We conduct performance evaluation on the two metrics shown above. We show the best validation accuracies in bold. The logistic regression validation accuracy is significantly below those of the more complex models we test.  While as the model complexity increases, the train accuracy also does, the validation accuracy seems to approach a limit of around $92.2\%$. For SVM the kernel choice is shown in parenthesis, the polynomial kernel was of degree 8.}
    \label{tab:model_comparison3}
\end{table}

\begin{figure}
        \centering
        \includegraphics[width=0.50\textwidth]{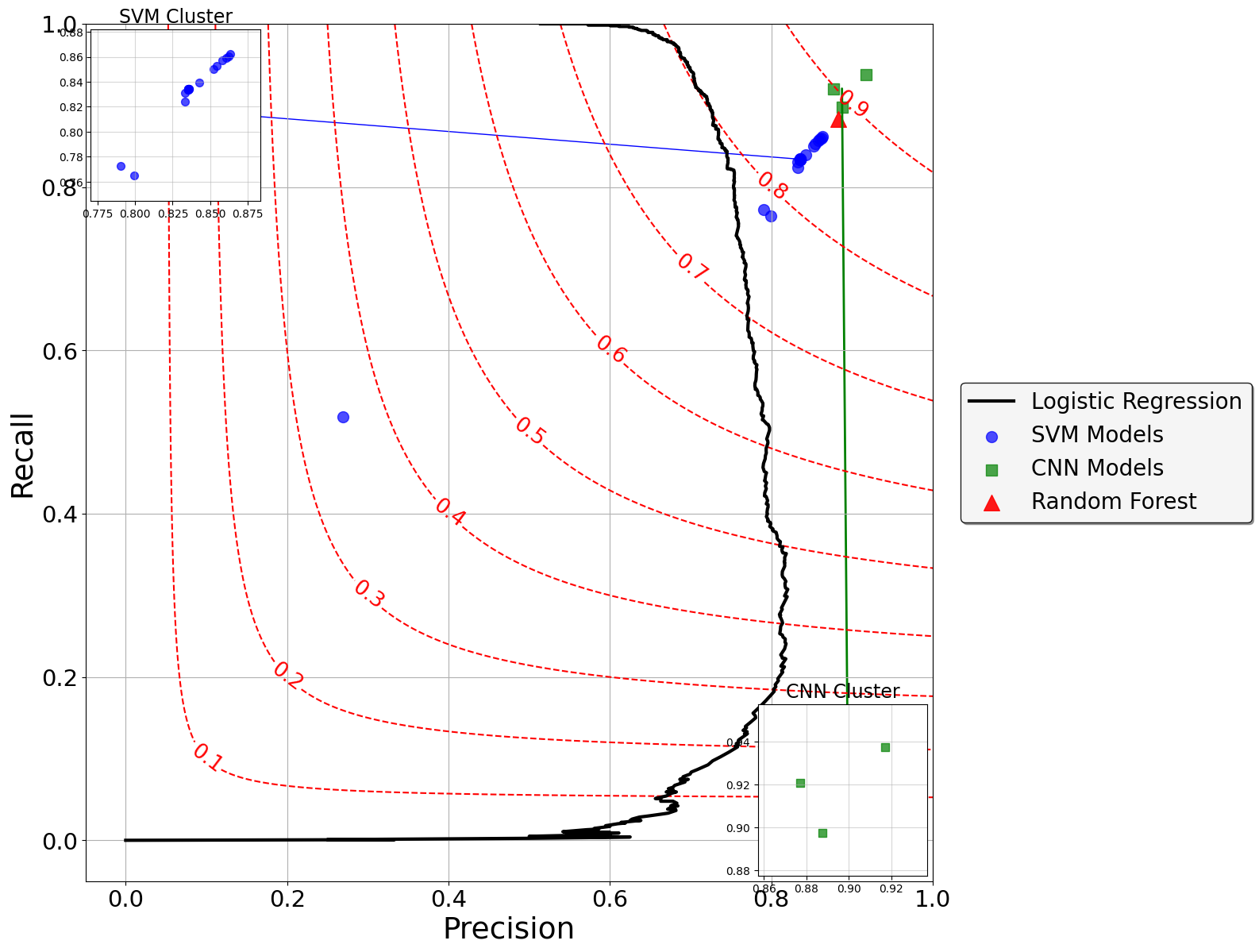}
        \caption{Here we show the trade-off between precision and recall for all the models we examine.  The logistic regression is computed for various values of the cut-off threshold and is shown as a black line.  The SVM, CNN and Random forest model are show as coloured points.  We also overlay contours of the $F_1$ sore given by equation \ref{eq(4)}.  We can see that out of the models we test, the CNN $M_{3}$ model maximizes the $F_1$ score and has the best precision-recall trade-off, though all models we examine perform moderately well.}
        \label{fig:pr-f1 curve}
    \end{figure}

    In Figure \ref{fig:pr-f1 curve}, we see the trade-off between precision and recall for each of the models. We deal with a solid curve for logistic regression, instead of individual data points since the logistic regression is evaluated on different thresholds, as explained in section \ref{logisticsectionref}, which leads to a continuous curve of data points for the model. On the contrary, for the other models, each one is represented by a single points, with SVMs and CNNs, being represented by more than one due to the multiple variations in the model architecture, owing to hyper-parameter tuning. Other than the logistic regression, our models lie near the $P=R$ line. 
    
    Therefore, precision is equal to recall ($P = R$). By looking at the figure, we can see that the CNN $M_{3}$ architecture outperforms all the other models since it is located at the highest possible $F_1$ score. Following that are the other CNN models, then random forest and SVM architectures and finally the logistic curve, which sets a baseline, as nearly all the other models perform better than the regression. Within the CNNs, Model $M_{3}$ performs the best with its complex architecture, while Model $M_{1}$ performs relatively worse with the simpler architecture, suggesting it might not be ideal for this kind of problem. 
    
    As mentioned in section \ref{cnn_explanation}, $M_{3}$ is an improvisation on $M_{2}$ which uses the \texttt{ReLU} activation function. We observe an increase in general performance compared to $M_{2}$. For the SVM cluster, we change the type of kernel (\texttt{Linear}, \texttt{RBF} or \texttt{Poly}) and the number of folds made over the dataset for each cross-validation pass. We conclude that the polynomial kernel results in the best precision-recall trade-off in the cluster, with the RBF kernel, while $9$ folds over the cross-validation split performs moderately good as well. The large number of points on the upper right are due to the iterative process which is carried over the 2 parameters: \texttt{num\_folds} and \texttt{Poly} kernel. The isolated point is for the 0 degree polynomial which severely inhibits the ability of the model to find an optimal hyperplane across the datapoints, it is the only model we test to be outperformed by the logistic regression.  Overall, we present a range of models that perform significantly better than a simple logistic approach. We show that even without time resolved light curve data, we can capture up to $\approx 90\%$ of the information for a transit signal, using only basic computational resources.
    
\subsection{Model Accuracy as a function of Flags}

\begin{figure}
        \centering
        \includegraphics[width=0.48\textwidth]{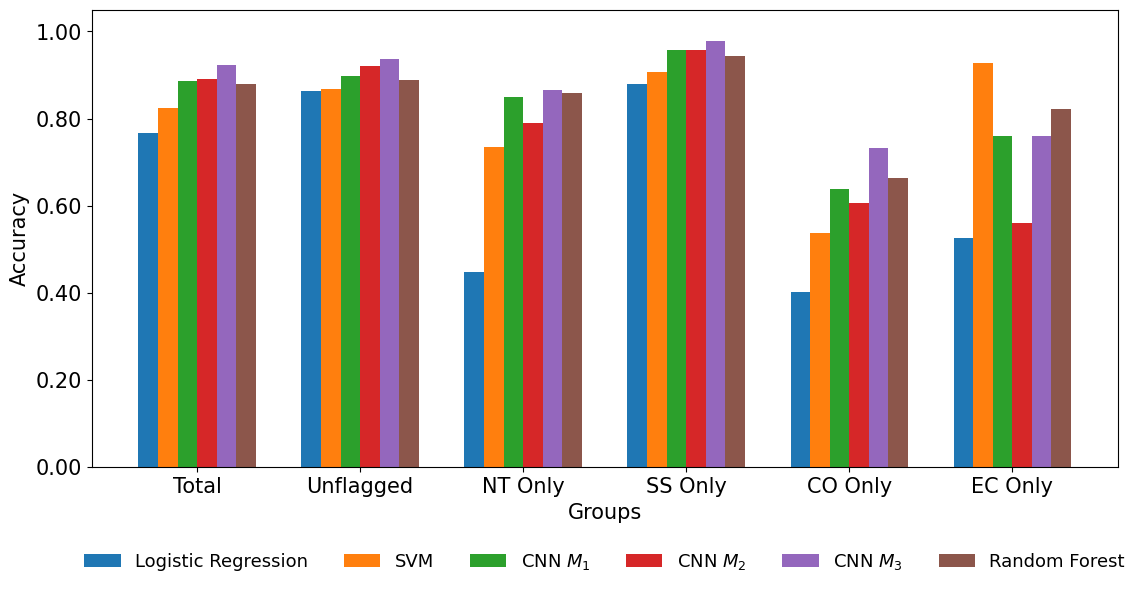}
        \caption{Model accuracy as a function of flag parameters (not transit (NT), stellar eclipse (SS), centroid offset (CO), ephemeris match contamination (EC)).}
        \label{fig:fig_last}
    \end{figure}

    We show our sub-classes results breakdown in figure \ref{fig:fig_last} in which we show the accuracy of all the models as a function of the flags on the test set. We consider six cases. The first set is the full test set with no regard for flags. The second set contains only objects that have no flags.  The remaining four sets consist of only objects with a given flag. 
    Since objects with flags are highly likely to be false positives, the $F_1$ score a poor metric to compare the models. Therefore we report the accuracy for each model ($A = \frac{TP+TN}{TP+NP+FP+FN}$) over each set in figure \ref{fig:fig_last}. We notice that most of the models perform well, to similar upper limit discussed in section \ref{sec:final_results}. We can also see that the best performance is relative to the stellar eclipse flag, while the poorer performances are related to the \texttt{CO} and \texttt{EC}  flag, for all the models. 

    For random forests and CNN model $M_{1}$, the un-flagged accuracy is very similar to that of the test set as a whole. However, there are some difference between their respective accuracies, when parametrised by the flags. For example, we can say that our CNN models seem to be slightly  more general model than Random Forest since they has more correct predictions relative to the total testing set for 3 out of the 4 flags (not transit, stellar eclipse, ephemeris match contamination). For centroid offset flag, both models have similar accuracies, but all models struggle to make accurate positions on systems with these offsets based only on the derived planetary parameters. Therefore, our CNN models might have greater generalisability over the Random Forest, if applied on other datasets such as TESS or SuperWASP. 

    Another key finding we can see is that for all the models except the logistic regression, the accuracy as a function of the \texttt{NT} flag is greater than a threshold; in this case, $A(\texttt{NT}) > 0.75$. Finally, we can say that the models' accuracies on the subset of data with the \texttt{NT} flag is comparable to that on the total and unflagged sets - except again for the logistic regression.   The accuracy as a function of the stellar eclipse flag exceeds that of the unflagged or total dataset, suggesting all our models are particularly good at discerning false positives caused by these effects. This over performance can be up to $14.77\%$ depending on the model in question. 
    
    For the flag categories, we can safely conclude that our derived planetary parameters don't contain sufficient information to learn about the ephemeris contamination and centroid offset flags. Ephemeris contamination, in particular, is complicated and subject to specific models and architectures. We find that, uniquely, SVM outperforms our more complex CNN and Random Forest models. Taken together, figure \ref{fig:fig_last} suggests that lightweight models provide good accuracies both for the dataset as a whole while also being able to discriminate against several common sources of false positives.  This raises the possibility that these lightweight models could be utilised for other missions such as TESS, with the advantage of achieving a much faster computations and run speed for the models, while sacrificing only a little accuracy.

\section{Conclusion}
We present a new approach to exoplanet detection by evaluating and optimizing multiple Machine Learning (ML) methods on the archival Kepler dataset. We show that simple ML models, trained on derived planetary parameters, instead of direct light curves, can achieve up to $\approx 90\%$ accuracies in assessing the status of an observation. This suggests that lightweight and rapid algorithms for exoplanet classification can be developed at the cost of only a little accuracy. These methods may prove invaluable in the future for other large datasets that may require quick, computation-efficient classification of the vetting status of exoplanets.
\section{Data Availability}
    This paper makes use of the Kepler Dataset from the NASA Exoplanet Archive. All of the model architectures used in this are available at \url{https://github.com/AyanBinRafaih/Exoplanet-For-MLFP}. 

\section*{Acknowledgments}
This work was supported by the French government through the France 2030 investment plan managed by the National Research Agency (ANR), as part of the Initiative of Excellence of Université Côte d’Azur under reference number ANR-15-IDEX-01.

\bibliographystyle{mnras}
\bibliography{refs}

@ARTICLE{2002ApJ...580L.171M,
       author = {{Mandel}, Kaisey and {Agol}, Eric},
        title = "{Analytic Light Curves for Planetary Transit Searches}",
      journal = {\apjl},
         year = 2002,
        month = dec,
       volume = {580},
       number = {2},
        pages = {L171-L175},
          doi = {10.1086/345520},
archivePrefix = {arXiv},
       eprint = {astro-ph/0210099},
 primaryClass = {astro-ph},
       adsurl = {https://ui.adsabs.harvard.edu/abs/2002ApJ...580L.171M}
}

@ARTICLE{Wolszczan_1992,
       author = {{Wolszczan}, A. and {Frail}, D.~A.},
        title = "{A planetary system around the millisecond pulsar PSR1257 + 12}",
      journal = {\nat},
         year = 1992,
        month = jan,
       volume = {355},
       number = {6356},
        pages = {145-147},
          doi = {10.1038/355145a0},
       adsurl = {https://ui.adsabs.harvard.edu/abs/1992Natur.355..145W}
}

@ARTICLE{Mayor_1995,
       author = {{Mayor}, Michel and {Queloz}, Didier},
        title = "{A Jupiter-mass companion to a solar-type star}",
      journal = {\nat},
         year = 1995,
        month = nov,
       volume = {378},
       number = {6555},
        pages = {355-359},
          doi = {10.1038/378355a0},
       adsurl = {https://ui.adsabs.harvard.edu/abs/1995Natur.378..355M}
}

@ARTICLE{Charbonneau_2000,
       author = {{Charbonneau}, David and {Brown}, Timothy M. and {Latham}, David W. and {Mayor}, Michel},
        title = "{Detection of Planetary Transits Across a Sun-like Star}",
      journal = {\apjl},
         year = 2000,
        month = jan,
       volume = {529},
       number = {1},
        pages = {L45-L48},
          doi = {10.1086/312457},
archivePrefix = {arXiv},
       eprint = {astro-ph/9911436},
 primaryClass = {astro-ph},
       adsurl = {https://ui.adsabs.harvard.edu/abs/2000ApJ...529L..45C}
}

@ARTICLE{Ricker_2015,
       author = {{Ricker}, George R. and {Winn}, Joshua N. and {Vanderspek}, Roland and {Latham}, David W. and {Bakos}, G{\'a}sp{\'a}r {\'A}. and {Bean}, Jacob L. and {Berta-Thompson}, Zachory K. and {Brown}, Timothy M. and {Buchhave}, Lars and {Butler}, Nathaniel R. and {Butler}, R. Paul and {Chaplin}, William J. and {Charbonneau}, David and {Christensen-Dalsgaard}, J{\o}rgen and {Clampin}, Mark and {Deming}, Drake and {Doty}, John and {De Lee}, Nathan and {Dressing}, Courtney and {Dunham}, Edward W. and {Endl}, Michael and {Fressin}, Francois and {Ge}, Jian and {Henning}, Thomas and {Holman}, Matthew J. and {Howard}, Andrew W. and {Ida}, Shigeru and {Jenkins}, Jon M. and {Jernigan}, Garrett and {Johnson}, John Asher and {Kaltenegger}, Lisa and {Kawai}, Nobuyuki and {Kjeldsen}, Hans and {Laughlin}, Gregory and {Levine}, Alan M. and {Lin}, Douglas and {Lissauer}, Jack J. and {MacQueen}, Phillip and {Marcy}, Geoffrey and {McCullough}, Peter R. and {Morton}, Timothy D. and {Narita}, Norio and {Paegert}, Martin and {Palle}, Enric and {Pepe}, Francesco and {Pepper}, Joshua and {Quirrenbach}, Andreas and {Rinehart}, Stephen A. and {Sasselov}, Dimitar and {Sato}, Bun'ei and {Seager}, Sara and {Sozzetti}, Alessandro and {Stassun}, Keivan G. and {Sullivan}, Peter and {Szentgyorgyi}, Andrew and {Torres}, Guillermo and {Udry}, Stephane and {Villasenor}, Joel},
        title = "{Transiting Exoplanet Survey Satellite (TESS)}",
      journal = {Journal of Astronomical Telescopes, Instruments, and Systems},
         year = 2015,
        month = jan,
       volume = {1},
          eid = {014003},
        pages = {014003},
          doi = {10.1117/1.JATIS.1.1.014003},
       adsurl = {https://ui.adsabs.harvard.edu/abs/2015JATIS...1a4003R}
}

@ARTICLE{Kovacs_2002,
       author = {{Kov{\'a}cs}, G. and {Zucker}, S. and {Mazeh}, T.},
        title = "{A box-fitting algorithm in the search for periodic transits}",
      journal = {\aap},
         year = 2002,
        month = aug,
       volume = {391},
        pages = {369-377},
          doi = {10.1051/0004-6361:20020802},
archivePrefix = {arXiv},
       eprint = {astro-ph/0206099},
 primaryClass = {astro-ph},
       adsurl = {https://ui.adsabs.harvard.edu/abs/2002A&A...391..369K}
}

@ARTICLE{Hippke_2019,
       author = {{Hippke}, Michael and {Heller}, Ren{\'e}},
        title = "{Optimized transit detection algorithm to search for periodic transits of small planets}",
      journal = {\aap},
         year = 2019,
        month = mar,
       volume = {623},
          eid = {A39},
        pages = {A39},
          doi = {10.1051/0004-6361/201834672},
archivePrefix = {arXiv},
       eprint = {1901.02015},
 primaryClass = {astro-ph.EP},
       adsurl = {https://ui.adsabs.harvard.edu/abs/2019A&A...623A..39H}
}

@ARTICLE{Jehin_2011,
       author = {{Jehin}, E. and {Gillon}, M. and {Queloz}, D. and {Magain}, P. and {Manfroid}, J. and {Chantry}, V. and {Lendl}, M. and {Hutsem{\'e}kers}, D. and {Udry}, S.},
        title = "{TRAPPIST: TRAnsiting Planets and PlanetesImals Small Telescope}",
      journal = {The Messenger},
         year = 2011,
        month = sep,
       volume = {145},
        pages = {2-6},
       adsurl = {https://ui.adsabs.harvard.edu/abs/2011Msngr.145....2J}
}

@ARTICLE{Pepper_2007,
       author = {{Pepper}, Joshua and {Pogge}, Richard W. and {DePoy}, D.~L. and {Marshall}, J.~L. and {Stanek}, K.~Z. and {Stutz}, Amelia M. and {Poindexter}, Shawn and {Siverd}, Robert and {O'Brien}, Thomas P. and {Trueblood}, Mark and {Trueblood}, Patricia},
        title = "{The Kilodegree Extremely Little Telescope (KELT): A Small Robotic Telescope for Large-Area Synoptic Surveys}",
      journal = {\pasp},
         year = 2007,
        month = aug,
       volume = {119},
       number = {858},
        pages = {923-935},
          doi = {10.1086/521836},
archivePrefix = {arXiv},
       eprint = {0704.0460},
 primaryClass = {astro-ph},
       adsurl = {https://ui.adsabs.harvard.edu/abs/2007PASP..119..923P}
}

@ARTICLE{Butters_2010,
       author = {{Butters}, O.~W. and {West}, R.~G. and {Anderson}, D.~R. and {Collier Cameron}, A. and {Clarkson}, W.~I. and {Enoch}, B. and {Haswell}, C.~A. and {Hellier}, C. and {Horne}, K. and {Joshi}, Y. and {Kane}, S.~R. and {Lister}, T.~A. and {Maxted}, P.~F.~L. and {Parley}, N. and {Pollacco}, D. and {Smalley}, B. and {Street}, R.~A. and {Todd}, I. and {Wheatley}, P.~J. and {Wilson}, D.~M.},
        title = "{The first WASP public data release}",
      journal = {\aap},
         year = 2010,
        month = sep,
       volume = {520},
          eid = {L10},
        pages = {L10},
          doi = {10.1051/0004-6361/201015655},
archivePrefix = {arXiv},
       eprint = {1009.5306},
 primaryClass = {astro-ph.EP},
       adsurl = {https://ui.adsabs.harvard.edu/abs/2010A&A...520L..10B}
}

@ARTICLE{Wheatley_2018,
       author = {{Wheatley}, Peter J. and {West}, Richard G. and {Goad}, Michael R. and {Jenkins}, James S. and {Pollacco}, Don L. and {Queloz}, Didier and {Rauer}, Heike and {Udry}, St{\'e}phane and {Watson}, Christopher A. and {Chazelas}, Bruno and {Eigm{\"u}ller}, Philipp and {Lambert}, Gregory and {Genolet}, Ludovic and {McCormac}, James and {Walker}, Simon and {Armstrong}, David J. and {Bayliss}, Daniel and {Bento}, Joao and {Bouchy}, Fran{\c{c}}ois and {Burleigh}, Matthew R. and {Cabrera}, Juan and {Casewell}, Sarah L. and {Chaushev}, Alexander and {Chote}, Paul and {Csizmadia}, Szil{\'a}rd and {Erikson}, Anders and {Faedi}, Francesca and {Foxell}, Emma and {G{\"a}nsicke}, Boris T. and {Gillen}, Edward and {Grange}, Andrew and {G{\"u}nther}, Maximilian N. and {Hodgkin}, Simon T. and {Jackman}, James and {Jord{\'a}n}, Andr{\'e}s and {Louden}, Tom and {Metrailler}, Lionel and {Moyano}, Maximiliano and {Nielsen}, Louise D. and {Osborn}, Hugh P. and {Poppenhaeger}, Katja and {Raddi}, Roberto and {Raynard}, Liam and {Smith}, Alexis M.~S. and {Soto}, Maritza and {Titz-Weider}, Ruth},
        title = "{The Next Generation Transit Survey (NGTS)}",
      journal = {\mnras},
         year = 2018,
        month = apr,
       volume = {475},
       number = {4},
        pages = {4476-4493},
          doi = {10.1093/mnras/stx2836},
archivePrefix = {arXiv},
       eprint = {1710.11100},
 primaryClass = {astro-ph.EP},
       adsurl = {https://ui.adsabs.harvard.edu/abs/2018MNRAS.475.4476W}
}

@ARTICLE{Morton_2012,
       author = {{Morton}, Timothy D.},
        title = "{An Efficient Automated Validation Procedure for Exoplanet Transit Candidates}",
      journal = {\apj},
         year = 2012,
        month = dec,
       volume = {761},
       number = {1},
          eid = {6},
        pages = {6},
          doi = {10.1088/0004-637X/761/1/6},
archivePrefix = {arXiv},
       eprint = {1206.1568},
 primaryClass = {astro-ph.EP},
       adsurl = {https://ui.adsabs.harvard.edu/abs/2012ApJ...761....6M}
}

@software{Morton_2015,
       author = {{Morton}, Timothy D.},
        title = "{VESPA: False positive probabilities calculator}",
 howpublished = {Astrophysics Source Code Library, record ascl:1503.011},
         year = 2015,
        month = mar,
          eid = {ascl:1503.011},
       adsurl = {https://ui.adsabs.harvard.edu/abs/2015ascl.soft03011M}
}

@ARTICLE{Giacalone_2021,
       author = {{Giacalone}, Steven and {Dressing}, Courtney D. and {Jensen}, Eric L.~N. and {Collins}, Karen A. and {Ricker}, George R. and {Vanderspek}, Roland and {Seager}, S. and {Winn}, Joshua N. and {Jenkins}, Jon M. and {Barclay}, Thomas and {Barkaoui}, Khalid and {Cadieux}, Charles and {Charbonneau}, David and {Collins}, Kevin I. and {Conti}, Dennis M. and {Doyon}, Ren{\'e} and {Evans}, Phil and {Ghachoui}, Mourad and {Gillon}, Micha{\"e}l and {Guerrero}, Natalia M. and {Hart}, Rhodes and {Jehin}, Emmanu{\"e}l and {Kielkopf}, John F. and {McLean}, Brian and {Murgas}, Felipe and {Palle}, Enric and {Parviainen}, Hannu and {Pozuelos}, Francisco J. and {Relles}, Howard M. and {Shporer}, Avi and {Socia}, Quentin and {Stockdale}, Chris and {Tan}, Thiam-Guan and {Torres}, Guillermo and {Twicken}, Joseph D. and {Waalkes}, William C. and {Waite}, Ian A.},
        title = "{Vetting of 384 TESS Objects of Interest with TRICERATOPS and Statistical Validation of 12 Planet Candidates}",
      journal = {\aj},
         year = 2021,
        month = jan,
       volume = {161},
       number = {1},
          eid = {24},
        pages = {24},
          doi = {10.3847/1538-3881/abc6af},
archivePrefix = {arXiv},
       eprint = {2002.00691},
 primaryClass = {astro-ph.EP},
       adsurl = {https://ui.adsabs.harvard.edu/abs/2021AJ....161...24G}
}

@ARTICLE{Armstrong_2021,
       author = {{Armstrong}, David J. and {Gamper}, Jevgenij and {Damoulas}, Theodoros},
        title = "{Exoplanet validation with machine learning: 50 new validated Kepler planets}",
      journal = {\mnras},
         year = 2021,
        month = jul,
       volume = {504},
       number = {4},
        pages = {5327-5344},
          doi = {10.1093/mnras/staa2498},
archivePrefix = {arXiv},
       eprint = {2008.10516},
 primaryClass = {astro-ph.EP},
       adsurl = {https://ui.adsabs.harvard.edu/abs/2021MNRAS.504.5327A}
}

@ARTICLE{Koch_2010,
       author = {{Koch}, David G. and {Borucki}, William J. and {Basri}, Gibor and {Batalha}, Natalie M. and {Brown}, Timothy M. and {Caldwell}, Douglas and {Christensen-Dalsgaard}, J{\o}rgen and {Cochran}, William D. and {DeVore}, Edna and {Dunham}, Edward W. and {Gautier}, III, Thomas N. and {Geary}, John C. and {Gilliland}, Ronald L. and {Gould}, Alan and {Jenkins}, Jon and {Kondo}, Yoji and {Latham}, David W. and {Lissauer}, Jack J. and {Marcy}, Geoffrey and {Monet}, David and {Sasselov}, Dimitar and {Boss}, Alan and {Brownlee}, Donald and {Caldwell}, John and {Dupree}, Andrea K. and {Howell}, Steve B. and {Kjeldsen}, Hans and {Meibom}, S{\o}ren and {Morrison}, David and {Owen}, Tobias and {Reitsema}, Harold and {Tarter}, Jill and {Bryson}, Stephen T. and {Dotson}, Jessie L. and {Gazis}, Paul and {Haas}, Michael R. and {Kolodziejczak}, Jeffrey and {Rowe}, Jason F. and {Van Cleve}, Jeffrey E. and {Allen}, Christopher and {Chandrasekaran}, Hema and {Clarke}, Bruce D. and {Li}, Jie and {Quintana}, Elisa V. and {Tenenbaum}, Peter and {Twicken}, Joseph D. and {Wu}, Hayley},
        title = "{Kepler Mission Design, Realized Photometric Performance, and Early Science}",
      journal = {\apjl},
         year = 2010,
        month = apr,
       volume = {713},
       number = {2},
        pages = {L79-L86},
          doi = {10.1088/2041-8205/713/2/L79},
archivePrefix = {arXiv},
       eprint = {1001.0268},
 primaryClass = {astro-ph.EP},
       adsurl = {https://ui.adsabs.harvard.edu/abs/2010ApJ...713L..79K}
}

@software{2020ascl.soft12006C,
       author = {{Coughlin}, Jeffrey L.},
        title = "{Robovetter: Automatic vetting of Threshold Crossing Events (TCEs)}",
 howpublished = {Astrophysics Source Code Library, record ascl:2012.006},
         year = 2020,
        month = dec,
          eid = {ascl:2012.006},
       adsurl = {https://ui.adsabs.harvard.edu/abs/2020ascl.soft12006C}
}

@ARTICLE{2018ApJS..235...38T,
       author = {{Thompson}, Susan E. and {Coughlin}, Jeffrey L. and {Hoffman}, Kelsey and {Mullally}, Fergal and {Christiansen}, Jessie L. and {Burke}, Christopher J. and {Bryson}, Steve and {Batalha}, Natalie and {Haas}, Michael R. and {Catanzarite}, Joseph and {Rowe}, Jason F. and {Barentsen}, Geert and {Caldwell}, Douglas A. and {Clarke}, Bruce D. and {Jenkins}, Jon M. and {Li}, Jie and {Latham}, David W. and {Lissauer}, Jack J. and {Mathur}, Savita and {Morris}, Robert L. and {Seader}, Shawn E. and {Smith}, Jeffrey C. and {Klaus}, Todd C. and {Twicken}, Joseph D. and {Van Cleve}, Jeffrey E. and {Wohler}, Bill and {Akeson}, Rachel and {Ciardi}, David R. and {Cochran}, William D. and {Henze}, Christopher E. and {Howell}, Steve B. and {Huber}, Daniel and {Pr{\v{s}}a}, Andrej and {Ram{\'\i}rez}, Solange V. and {Morton}, Timothy D. and {Barclay}, Thomas and {Campbell}, Jennifer R. and {Chaplin}, William J. and {Charbonneau}, David and {Christensen-Dalsgaard}, J{\o}rgen and {Dotson}, Jessie L. and {Doyle}, Laurance and {Dunham}, Edward W. and {Dupree}, Andrea K. and {Ford}, Eric B. and {Geary}, John C. and {Girouard}, Forrest R. and {Isaacson}, Howard and {Kjeldsen}, Hans and {Quintana}, Elisa V. and {Ragozzine}, Darin and {Shabram}, Megan and {Shporer}, Avi and {Silva Aguirre}, Victor and {Steffen}, Jason H. and {Still}, Martin and {Tenenbaum}, Peter and {Welsh}, William F. and {Wolfgang}, Angie and {Zamudio}, Khadeejah A. and {Koch}, David G. and {Borucki}, William J.},
        title = "{Planetary Candidates Observed by Kepler. VIII. A Fully Automated Catalog with Measured Completeness and Reliability Based on Data Release 25}",
      journal = {\apjs},
         year = 2018,
        month = apr,
       volume = {235},
       number = {2},
          eid = {38},
        pages = {38},
          doi = {10.3847/1538-4365/aab4f9},
archivePrefix = {arXiv},
       eprint = {1710.06758},
 primaryClass = {astro-ph.EP},
       adsurl = {https://ui.adsabs.harvard.edu/abs/2018ApJS..235...38T}
}

@MISC{2017ksci.rept...22C,
       author = {{Coughlin}, Jeffrey L.},
        title = "{Planet Detection Metrics: Robovetter Completeness and Effectiveness for Data Release 25}",
 howpublished = {Kepler Science Document KSCI-19114-002, id. 22. Edited by Natalie Batalha and Michael R. Haas},
         year = 2017,
        month = aug,
          eid = {22},
        pages = {22},
       adsurl = {https://ui.adsabs.harvard.edu/abs/2017ksci.rept...22C}
}

@misc{kingma2017adammethodstochasticoptimization,
      title={Adam: A Method for Stochastic Optimization}, 
      author={Diederik P. Kingma and Jimmy Ba},
      year={2017},
      eprint={1412.6980},
      archivePrefix={arXiv},
      primaryClass={cs.LG},
      url={https://arxiv.org/abs/1412.6980}, 
}

@ARTICLE{TardugnoPoleo_2024,
       author = {{Tardugno Poleo}, Valentina and {Eisner}, Nora and {Hogg}, David W.},
        title = "{NotPlaNET: Removing False Positives from Planet Hunters TESS with Machine Learning}",
      journal = {\aj},
         year = 2024,
        month = sep,
       volume = {168},
       number = {3},
          eid = {100},
        pages = {100},
          doi = {10.3847/1538-3881/ad5f29},
archivePrefix = {arXiv},
       eprint = {2405.18278},
 primaryClass = {astro-ph.EP},
       adsurl = {https://ui.adsabs.harvard.edu/abs/2024AJ....168..100T}
}

@ARTICLE{Malik_2022,
       author = {{Malik}, Shreshth A. and {Eisner}, Nora L. and {Lintott}, Chris J. and {Gal}, Yarin},
        title = "{Discovering Long-period Exoplanets using Deep Learning with Citizen Science Labels}",
      journal = {arXiv e-prints},
         year = 2022,
        month = nov,
          eid = {arXiv:2211.06903},
        pages = {arXiv:2211.06903},
          doi = {10.48550/arXiv.2211.06903},
archivePrefix = {arXiv},
       eprint = {2211.06903},
 primaryClass = {astro-ph.EP},
       adsurl = {https://ui.adsabs.harvard.edu/abs/2022arXiv221106903M}
}

@ARTICLE{2003ApJ...593L.125B,
       author = {{Brown}, Timothy M.},
        title = "{Expected Detection and False Alarm Rates for Transiting Jovian Planets}",
      journal = {\apjl},
         year = 2003,
        month = aug,
       volume = {593},
       number = {2},
        pages = {L125-L128},
          doi = {10.1086/378310},
archivePrefix = {arXiv},
       eprint = {astro-ph/0307256},
 primaryClass = {astro-ph},
       adsurl = {https://ui.adsabs.harvard.edu/abs/2003ApJ...593L.125B}
}

@ARTICLE{2006ApJ...651L..61O,
       author = {{O'Donovan}, Francis T. and {Charbonneau}, David and {Mandushev}, Georgi and {Dunham}, Edward W. and {Latham}, David W. and {Torres}, Guillermo and {Sozzetti}, Alessandro and {Brown}, Timothy M. and {Trauger}, John T. and {Belmonte}, Juan A. and {Rabus}, Markus and {Almenara}, Jos{\'e} M. and {Alonso}, Roi and {Deeg}, Hans J. and {Esquerdo}, Gilbert A. and {Falco}, Emilio E. and {Hillenbrand}, Lynne A. and {Roussanova}, Anna and {Stefanik}, Robert P. and {Winn}, Joshua N.},
        title = "{TrES-2: The First Transiting Planet in the Kepler Field}",
      journal = {\apjl},
         year = 2006,
        month = nov,
       volume = {651},
       number = {1},
        pages = {L61-L64},
          doi = {10.1086/509123},
archivePrefix = {arXiv},
       eprint = {astro-ph/0609335},
 primaryClass = {astro-ph},
       adsurl = {https://ui.adsabs.harvard.edu/abs/2006ApJ...651L..61O}
}

@article{scikit-learn,
  title={Scikit-learn: Machine Learning in {P}ython},
  author={Pedregosa, F. and Varoquaux, G. and Gramfort, A. and Michel, V.
          and Thirion, B. and Grisel, O. and Blondel, M. and Prettenhofer, P.
          and Weiss, R. and Dubourg, V. and Vanderplas, J. and Passos, A. and
          Cournapeau, D. and Brucher, M. and Perrot, M. and Duchesnay, E.},
  journal={Journal of Machine Learning Research},
  volume={12},
  pages={2825--2830},
  year={2011}
}

@misc{abadi2016tensorflowlargescalemachinelearning,
      title={TensorFlow: A system for large-scale machine learning}, 
      author={Martín Abadi and Paul Barham and Jianmin Chen and Zhifeng Chen and Andy Davis and Jeffrey Dean and Matthieu Devin and Sanjay Ghemawat and Geoffrey Irving and Michael Isard and Manjunath Kudlur and Josh Levenberg and Rajat Monga and Sherry Moore and Derek G. Murray and Benoit Steiner and Paul Tucker and Vijay Vasudevan and Pete Warden and Martin Wicke and Yuan Yu and Xiaoqiang Zheng},
      year={2016},
      eprint={1605.08695},
      archivePrefix={arXiv},
      primaryClass={cs.DC},
      url={https://arxiv.org/abs/1605.08695}, 
}

@ARTICLE{Borucki_2010,
       author = {{Borucki}, William J. and {Koch}, David and {Basri}, Gibor and {Batalha}, Natalie and {Brown}, Timothy and {Caldwell}, Douglas and {Caldwell}, John and {Christensen-Dalsgaard}, J{\o}rgen and {Cochran}, William D. and {DeVore}, Edna and {Dunham}, Edward W. and {Dupree}, Andrea K. and {Gautier}, Thomas N. and {Geary}, John C. and {Gilliland}, Ronald and {Gould}, Alan and {Howell}, Steve B. and {Jenkins}, Jon M. and {Kondo}, Yoji and {Latham}, David W. and {Marcy}, Geoffrey W. and {Meibom}, S{\o}ren and {Kjeldsen}, Hans and {Lissauer}, Jack J. and {Monet}, David G. and {Morrison}, David and {Sasselov}, Dimitar and {Tarter}, Jill and {Boss}, Alan and {Brownlee}, Don and {Owen}, Toby and {Buzasi}, Derek and {Charbonneau}, David and {Doyle}, Laurance and {Fortney}, Jonathan and {Ford}, Eric B. and {Holman}, Matthew J. and {Seager}, Sara and {Steffen}, Jason H. and {Welsh}, William F. and {Rowe}, Jason and {Anderson}, Howard and {Buchhave}, Lars and {Ciardi}, David and {Walkowicz}, Lucianne and {Sherry}, William and {Horch}, Elliott and {Isaacson}, Howard and {Everett}, Mark E. and {Fischer}, Debra and {Torres}, Guillermo and {Johnson}, John Asher and {Endl}, Michael and {MacQueen}, Phillip and {Bryson}, Stephen T. and {Dotson}, Jessie and {Haas}, Michael and {Kolodziejczak}, Jeffrey and {Van Cleve}, Jeffrey and {Chandrasekaran}, Hema and {Twicken}, Joseph D. and {Quintana}, Elisa V. and {Clarke}, Bruce D. and {Allen}, Christopher and {Li}, Jie and {Wu}, Haley and {Tenenbaum}, Peter and {Verner}, Ekaterina and {Bruhweiler}, Frederick and {Barnes}, Jason and {Prsa}, Andrej},
        title = "{Kepler Planet-Detection Mission: Introduction and First Results}",
      journal = {Science},
         year = 2010,
        month = feb,
       volume = {327},
       number = {5968},
        pages = {977},
          doi = {10.1126/science.1185402},
       adsurl = {https://ui.adsabs.harvard.edu/abs/2010Sci...327..977B}
}

@ARTICLE{Coughlin_2016,
       author = {{Coughlin}, Jeffrey L. and {Mullally}, F. and {Thompson}, Susan E. and {Rowe}, Jason F. and {Burke}, Christopher J. and {Latham}, David W. and {Batalha}, Natalie M. and {Ofir}, Aviv and {Quarles}, Billy L. and {Henze}, Christopher E. and {Wolfgang}, Angie and {Caldwell}, Douglas A. and {Bryson}, Stephen T. and {Shporer}, Avi and {Catanzarite}, Joseph and {Akeson}, Rachel and {Barclay}, Thomas and {Borucki}, William J. and {Boyajian}, Tabetha S. and {Campbell}, Jennifer R. and {Christiansen}, Jessie L. and {Girouard}, Forrest R. and {Haas}, Michael R. and {Howell}, Steve B. and {Huber}, Daniel and {Jenkins}, Jon M. and {Li}, Jie and {Patil-Sabale}, Anima and {Quintana}, Elisa V. and {Ramirez}, Solange and {Seader}, Shawn and {Smith}, Jeffrey C. and {Tenenbaum}, Peter and {Twicken}, Joseph D. and {Zamudio}, Khadeejah A.},
        title = "{Planetary Candidates Observed by Kepler. VII. The First Fully Uniform Catalog Based on the Entire 48-month Data Set (Q1-Q17 DR24)}",
      journal = {\apjs},
         year = 2016,
        month = may,
       volume = {224},
       number = {1},
          eid = {12},
        pages = {12},
          doi = {10.3847/0067-0049/224/1/12},
archivePrefix = {arXiv},
       eprint = {1512.06149},
 primaryClass = {astro-ph.EP},
       adsurl = {https://ui.adsabs.harvard.edu/abs/2016ApJS..224...12C}
}
\end{document}